\documentclass[aps,pra,reprint,a4paper,superscriptaddress,numbers,longbibliography,showpacs,showkeys]{revtex4-1}

\usepackage[pdftex]{hyperref,color,graphicx}
\usepackage{amsfonts,amssymb,amsmath}
\usepackage[separate-uncertainty=true]{siunitx}
\usepackage[english]{babel}
\usepackage[utf8]{inputenc}
\usepackage[T1]{fontenc}

\def\bra#1{\mathinner{\langle{#1}|}}
\def\ket#1{\mathinner{|{#1}\rangle}}
\def\braket#1{\mathinner{\langle{#1}\rangle}}

\newif\ifusebibfile
\usebibfiletrue
\newcommand{\figref}[2]{\hyperref[#1]{\ref{#1}(#2)}} 

\hypersetup{
    pdftitle = {Robustness of topologically protected edge states in quantum walk experiments with neutral atoms},
    pdfauthor = {Thorsten Groh, Stefan Brakhane, Wolfgang Alt, Dieter Meschede, Janos Asb\'oth, Andrea Alberti},
	colorlinks=true,linkcolor=blue,citecolor=blue,filecolor=blue,urlcolor=blue
  }

\newcommand{\figlab}[1]{{\textbf{{#1}}}} 
\newcommand{\vect}[1]{\boldsymbol{#1}} 
\newcommand{\op}[1]{{{#1}}} 
\newcommand{\imag}{{\mathrm{i}}} 
\newcommand{\e}{{\mathrm{e}}} 

\renewcommand\textemdash{\leavevmode\unskip\kern0.8pt\rule[0.19\baselineskip]{8pt}{0.4pt}\kern1pt\ignorespaces}

\newsavebox{\diamondSymBox}
\savebox{\diamondSymBox}{\includegraphics[width=5pt]{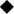}}
\newcommand{\diamondSym}{\usebox{\diamondSymBox}}

\newsavebox{\starSymBox}
\savebox{\starSymBox}{\includegraphics[width=5pt]{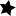}}
\newcommand{\starSym}{\usebox{\starSymBox}}

\begin{document}
\selectlanguage{english}

\title{Robustness of topologically protected edge states in quantum walk experiments \texorpdfstring{\\}{} with neutral atoms}

\author{Thorsten Groh}
\author{Stefan Brakhane}
\author{Wolfgang Alt}
\author{Dieter Meschede}
\affiliation{Institut für Angewandte Physik, Universität Bonn,
Wegelerstr.~8, D-53115 Bonn, Germany}
\author{Janos K. Asbóth}
\affiliation{Institute for Solid State Physics and Optics, Wigner Research
Centre for Physics, Hungarian Academy of Sciences, H-1525 Budapest P.O. Box 49,
Hungary}
\author{Andrea Alberti}
\email{alberti@iap.uni-bonn.de}
\affiliation{Institut für Angewandte Physik, Universität Bonn,
Wegelerstr.~8, D-53115 Bonn, Germany}
\date{\today}



\begin{abstract}
Discrete-time quantum walks allow Floquet topological insulator materials to be explored using  controllable systems such as ultracold atoms in optical lattices.
By numerical simulations, we study the robustness of topologically protected edge states in the presence of decoherence in one- and two-dimensional discrete-time quantum walks.
We also develop a simple analytical model quantifying the robustness of these edge states against either spin or spatial dephasing, predicting an exponential decay of the population of topologically protected edge states.
Moreover, we present an experimental proposal based on neutral atoms in spin-dependent optical lattices to realize spatial boundaries between distinct topological phases.
Our proposal relies on a new scheme to implement spin-dependent discrete shift operations in a two-dimensional optical lattice.
We analyze under realistic decoherence conditions the experimental feasibility of observing unidirectional, dissipationless transport of matter waves along boundaries separating distinct topological domains.
\end{abstract}

\maketitle

\section{Introduction}
Topological insulators are quantum materials behaving like an ordinary insulator in the bulk, and yet allowing, in two dimensions and above, matter waves to propagate along their boundaries through a discrete number of edge modes \cite{Hasan2010, bernevig2013topological}.
The distinguishing property of these materials is the existence of so-called topologically protected (TP) edge modes, which are robust against continuous deformations of the material's parameters including spatial disorder, providing the bulk remains insulating (i.e., no gap closing).
In one dimension (1D), a discrete number of TP edge states can exist in the presence of special symmetries (e.g., particle-hole symmetry in superconducting quantum wires), with their energy being exactly pinned to the midpoint of the energy gap.
In two dimensions (2D), the most notable example of a topological insulator is a two-dimensional electron gas in a high magnetic field, where the transverse conductance is found to be quantized in multiples of $e^2/h$ (integer quantum Hall effect, IQHE) \cite{Haug:1993}.
Over the years, this effect has been verified by experiments to one part in $10^9$ despite impurities and other imperfections, which unavoidably occur in actual physical samples \cite{vonKlitzing:2005}.
Its robustness is today well understood in terms of the topological structure of the Landau levels, which form well-separated energy bands \cite{Avron:2003}.

In general, the robustness of edge states in these insulating materials results from energy bands with nontrivial topological character.
Topologically nontrivial bands are often related to an obstruction to define the Bloch wave functions over the whole Brillouin zone using a single phase convention \cite{Fruchart:2013}.
This obstruction to a global choice of the gauge can be understood as resulting from a twist of the Bloch wave functions, much as the twist in the Möbius strip represents an obstruction to define an oriented surface.
The twists of the energy bands are quantified by topological invariants, which are integer quantum numbers assigned to each isolated band of the bulk.
These can be, for instance, winding numbers $\mathbb{Z}$ (e.g., for the Su-Schrieffer-Heeger model), or just $\mathbb{Z}_2$ numbers with two possible values denoting trivial and nontrivial topological phases (e.g., for particle-hole-symmetric quantum wires).
The characteristic of such invariants is that they are unchanged under a continuous modification of the system parameters, provided that the energy gap and the relevant symmetries are preserved.
In particular, two insulators are said to belong to different topological phases if the sum of the topological invariants of occupied bands are different \cite{kitaev2009periodic,Ryu2010}.

A topological argument with far-reaching physical implications, known as the \emph{bulk-boundary correspondence principle}, establishes a relation between the topological invariants and the  number of TP edge modes at the boundary between two topological phases \cite{Asboth:Book}.
Simply stated, it predicts that any spatial crossover region separating two bulks hosts a minimum number of edge modes given by the difference of the bulk invariants.
These modes are topologically protected as they cannot disappear by a continuous deformation of the system parameters, including a deformation of the boundary's shape.
In the IQHE, for instance, the number of current-carrying TP edge modes is equal to the sum of the Chern numbers of the Landau levels below the Fermi energy \footnote{The Chern numbers are the Topological invariants appropriate to classify time-independent two-dimensional topological insulators of non-interacting particles.}.

TP edge modes at the boundary of a 2D topological insulator are immune to Anderson localization.
Even if we allow for local disorder (of any amount in the region adjacent to the boundary), including shape irregularities, topological arguments predict that TP edge states maintain their metallic-like character notwithstanding the disorder\textemdash their wave functions being fully delocalized around the whole length of the insulator \cite{Asboth:Book}.
As a consequence, any wave packet formed by a superposition of TP edge states propagates coherently along the boundary, instead of being confined within some region by the disorder.
Moreover, transport along the boundary is virtually immune to backscattering too \cite{Buttiker:1988},
for the wave packet would need to tunnel to the opposite edge of the insulator material in order to couple to a counter-propagating edge mode \textemdash a process that is exponentially suppressed with the size of the sample.

Besides being interesting \emph{per se}, topological insulators have stimulated great interest for the possibility to exploit TP edge states for engineering ballistic electronic transport in dissipationless solid-state devices and for enabling topological protection of quantum information \cite{DasSarma:2015}.
In recent years, IQHE devices have attained an exquisite level of control, which enabled the demonstration of quantum devices such as an electronic Mach-Zehnder interferometer \cite{Ji:2003} and a two-electron Hong-Ou-Mandel-like interferometer \cite{Bocquillon:2013}.
However, these systems still require high magnetic fields on the order of $\SI{10}{\tesla}$ in order to make the energy gap between Landau levels (i.e., the cyclotron frequency) larger than cryogenic temperatures below $\SI{4}{\kelvin}$.
Larger gaps are obtained with high-mobility graphene IQHE devices, holding promise to operate at room temperature, though still requiring high magnetic fields \cite{Novoselov:2007}.
In a different approach, the quantum anomalous Hall effect avoids external magnetic fields by exploiting a ferromagnetic topological-insulator state induced by spontaneous magnetization, though demanding, in return, cryogenic temperatures well below both the Curie point and the magnetically induced energy gap \cite{Chang:2013,Katmis:2016}.
The discovery of quantum spin Hall effect in HgTe/CdTe quantum wells started the quest for topological insulators with large gap, and yet not relying on magnetic fields \cite{Konig:2007}.
However, the gap size of these novel materials still imposes, at least so far, cryogenic temperatures $<\SI{10}{\kelvin}$ to function \cite{Li:2015b}.

Topological insulator materials are challenging to synthesize, and only a few topological phases have hitherto been accessible with solid-state materials \cite{Ando2013}.
This has motivated the search for topological phases in non-electronic systems, which also allow implementing the same wave-mechanical principles underlying topological insulators.
Because of their high degree of control and flexibility, ultracold atoms trapped in an optical lattice are ideal systems to shed new light on the origin and dynamics of topological insulators.
In particular, these systems have enabled the direct measurement of the Berry-Zak phase \cite{Atala:2013} and Wilson lines \cite{Li:2016}, the realization of the Haldane model \cite{Jotzu:2014}, the measurements of the anomalous transverse velocity \cite{Aidelsburger:2014}, demonstration of the Thouless pump mechanism \cite{Lohse:2015,Nakajima:2016}, the realization of compacted artificial dimensions \cite{Mancini:2015,Stuhl:2015}, and the measurement of the Berry flux \cite{Duca:2014} as well as Berry curvature \cite{Flaschner:2016}.
Besides ultracold atom systems, TP edge modes have also been observed in microwave photonic crystals \cite{Wang:2009}, photonic quasicrystals \cite{Kraus:2012,Bandres:2016}, and even mechanical spring systems \cite{Kane2013,Susstrunk:2015}.

Discrete-time quantum walks (DTQWs) with trapped ultracold atoms \cite{Karski2009} offer a versatile and highly controlled platform for the experimental investigation of topological insulators.
We note that even a single atom coherently delocalized on a periodic potential is sufficient to simulate topology-induced transport phenomena, provided that the energy bands have a nontrivial topological structure.
In DTQW experiments, an ultracold atom trapped in an optical lattice undergoes a periodic sequence of internal rotations and spin-dependent translations.
This approach can be understood to fall under the more general class of Floquet topological insulators \textemdash systems that are periodically driven in time with a period $T$.
After an integer number of periods (i.e.\ steps), their quantum evolution is reproduced by an effective (Floquet) Hamiltonian that is topologically nontrivial \cite{cayssol2013floquet}.
Varying the protocol for the DTQW is a mean to engineer the effective Hamiltonian.
In this way, effective Hamiltonians from all universality classes of topological insulators \cite{kitaev2009periodic,Ryu2010} can be realized by quantum walks \cite{Kitagawa2010Expl}.

Floquet topological insulators are especially attractive for the possibility to control their topological properties via an external periodic drive \cite{Lindner:2011}, yet avoiding any external magnetic field.
An optical analogue of Floquet topological insulators was demonstrated using an array of evanescently coupled waveguides on a honeycomb lattice \cite{Rechtsman:2013}, with
the external periodic drive being effectively implemented by a helicoidal deformation of the waveguides.
DTQWs are well suited for creating TP edge modes, on the fly, by locally controlling the parameters of the external drive.
Furthermore, beyond simulating static topological insulators, DTQWs allow us to explore the richer topological structure inherent to Floquet systems, which is not entirely represented in the effective Hamiltonian, but instead rooted in the details of the quantum walk sequence.
For example, a one-dimensional quantum walk can host TP edge states between domains having the same effective Hamiltonian \cite{Asboth2012}.
Experimental evidence of this phenomenon was shown in a photonic DTQW setup, though using only a small number of steps \cite{kitagawa2012observation}.

In our laboratory we choose a single massive Cs atom with two long-lived hyperfine states as the quantum walker, which we coherently delocalized in optical lattices over ten and more lattice sites \cite{Alberti2014}.
However, quantum superposition states in such a large Hilbert space are always highly fragile because they are subject to decoherence and dephasing mechanisms arising from the openness of the quantum system.
In DTQWs decoherence leads to a quantum-to-classical transition of the walk evolution dominated by the dephasing process affecting the coherences in the coin degree of freedom, as we have shown previously \cite{Alberti2014}. 
It is generally accepted that disturbances with frequencies beyond the energy gap lead to the destruction of the TP edge states.
However, in most condensed matter systems, these effects are often suppressed by operating at cryogenic temperatures \cite{Martin:1990}.
In DTQWs, disturbances on the coin operation, as well as spin dephasing, effectively act with infinitely wide spectrum and therefore extend over the whole band gap, so that we expect the loss of protection in the long time limit.
In the 1D split-step walk, Obuse et al.~\cite{Obuse2011} has shown, that while topological protection is preserved under weak spatial disorder, temporal fluctuations of the coin angles destroy it.
However, a quantitative modeling of decoherence effects, which is essential for future experiments, is still missing.

In this paper, we study how environment-induced dephasing affects TP edge states in one- and two-dimensional quantum walk setups and how diffusive spreading has an impact on the existence and form of TP edge states in general.
Moreover, we formulate an experimental proposal under realistic conditions on how to observe ballistic transport of quantum walks using ultracold atoms in optical lattices.
 
The paper is structured as follows:
In Sec.~\ref{sec:TopPhases}, we introduce DTQW protocols in one and two dimensions and provide a short overview of their topological structure and corresponding TP edge states.
We discuss the arising edge phenomena and analyze their robustness under spatial deformations of the topological phase boundary.
In Sec.~\ref{sec:robustness}, we investigate how the shape and evolution of the edge states is affected under decoherence.
Furthermore, we give insight into the limits concerning the model of stroboscopic decoherence, which was employed in Ref.~\citenum{Alberti2014}.
The numerical simulations in this analysis are carried out using realistic experimental parameters, which are chosen based on the experimental proposal discussed in Sec.~\ref{sec:experiment}.
In Sec.~\ref{sec:experiment}, we present a new experimental scheme to realize a two-dimensional spin-dependent optical lattice, and discuss the experimental requirements to create spatial boundaries between topological phases as well as to observe TP edge states under realistic decoherence conditions. 

\section{Topological phases in discrete-time quantum walks}
\label{sec:TopPhases}

\subsection{The system}

We consider a particle with two internal spin states, labeled $s \in \{\uparrow, \downarrow\}$, that is positioned on a cubic lattice with lattice constant $a$.
We will specifically address the cases of $N = 1$ and $N = 2$ dimensions, which can be implemented in present experimental apparatuses, as explained in detail in Sec.~\ref{sec:experiment}.
We label the nodes of the $N$-dimensional cubic lattice with $\vect{x} = (x, y, \ldots) \in \mathbb{Z}^N$.
Thus, in the absence of decoherence, the quantum state of the walker after $n$ steps is a pure state $\ket{\psi_n}$, which comprises a superposition of the basis states $\ket{\vect{x},s}$.

The dynamics of the DTQW is defined by a sequence of unitary operations (\emph{protocol}), which can be of two types: the \emph{coin toss} operation and \emph{spin-dependent shift} operations.
The coin toss is realized by a unitary rotation of the spin state into superpositions of $\ket{\uparrow}$ and $\ket{\downarrow}$,
\begin{equation}
	\label{eq:coindefinition}
    \op{C}(\theta) = \sum_{\vect{x}} \ket{\vect{x}}\!\bra{\vect{x}} \otimes
    \e^{-\imag\hspace{0.3pt} \op{\sigma}_2\hspace{0.3pt}\theta / 2} ,    
\end{equation}
where $\op{\sigma}_i$ is the $i$-th Pauli matrix.
The {coin angle} $\theta$ determines the amount of rotation of the spin state and is, in general, a function of the lattice position $\vect{x}$, $\theta = \theta({\vect{x}})$.
The rotation axis does not depend on the position, instead, and is chosen to be along the $y$-direction of the Bloch sphere.
Note that different choices of the rotation axis in the $x$-$y$ plane are equivalent up to a unitary transformation of the spin basis vectors $\{\ket{\uparrow}, \ket{\downarrow}\}$.

Different choices of the rotation axis are equivalent to a unitary transformation of the spin basis vectors $\{\ket{\uparrow}, \ket{\downarrow}\}$.

The spin-dependent shift operation $\op{S}_d^s$ ($s \in \{\uparrow, \downarrow\}$, $d \in \{x,y\}$) is defined as
\begin{align}
    \op{S}_d^\uparrow &= \sum_{\vect{x}} \ket{\vect{x}+\vect{e}_d}\!\bra{\vect{x}}
    \otimes \ket{\uparrow}\!\bra{\uparrow} + \ket{\vect{x}}\!\bra{\vect{x}} 
    \otimes \ket{\downarrow}\!\bra{\downarrow} , \\
    \op{S}_d^\downarrow &= \sum_{\vect{x}} \ket{\vect{x}-\vect{e}_d}\!\bra{\vect{x}}
    \otimes \ket{\downarrow}\!\bra{\downarrow} + \ket{\vect{x}}\!\bra{\vect{x}}
    \otimes \ket{\uparrow}\!\bra{\uparrow} ,
\end{align}
where $\vect{e}_d$ denotes the unit lattice vector in the $d$-direction. 
$\op{S}_d^\uparrow$ ($\op{S}_d^\downarrow$) shifts the walker's spin up (down) component in the positive (negative) $\vect{e}_d$-direction by one lattice site, while the other spin component is unchanged. 

The evolution of a pure state $\ket{\psi_n}$ in time is described by a unitary walk operator $\op{W}$ applied periodically at discrete time steps $t=n\,T$, $n\in\mathbb{N}$:
\begin{equation}
    \ket{\psi_n} = \op{W}^n \ket{\psi_0} .
\end{equation}
Note that the quantum evolution of the walker is periodically driven in time with a Floquet period $T$, which is the duration of a single step.

In this work we focus on two DTQW protocols, which allow us to study the most relevant physical properties of topological phases of discrete-time quantum walks in one- and two-dimensions. In a one-dimensional (1D) lattice, we consider the so-called \emph{split-step walk} protocol defined in Ref.~\citenum{Kitagawa2010Expl} as
\begin{equation}
    \op{W}_\text{1D} = \op{S}_x^\downarrow\,\op{C}(\theta_2)\, 
                     \op{S}_x^\uparrow\, \op{C}(\theta_1) ,
    \label{eq:1d_protocol}
\end{equation}
consisting of two spin rotations separated by spin-dependent shifts in $x$-direction. 
In a two-dimensional (2D) lattice, we study the quantum walk defined by
\begin{equation}
    \op{W}_\text{2D} = 
    \op{S}_y^\downarrow\,\op{S}_y^\uparrow\, \op{C}(\theta_2)\, 
    \op{S}_x^\downarrow\,\op{S}_x^\uparrow\, \op{C}(\theta_1),
    \label{eq:2d_protocol}
\end{equation}
where after each coin operation both spin states are shifted in opposite directions \cite{Kitagawa2012}.
Note that the shift operators commute, $[\op{S}_d^\uparrow,\op{S}_d^\downarrow]=0$.

\begin{figure*}
    \includegraphics[width=1\textwidth]{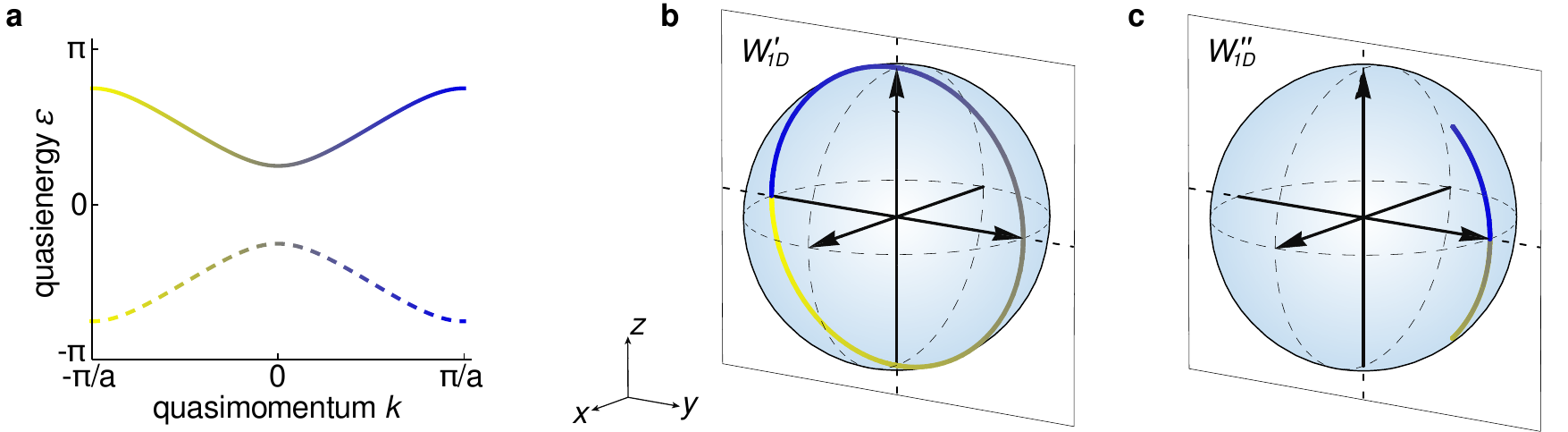}
    \caption{
    Topological twist in the 1D split-step quantum walk with $(\theta_1, \theta_2) = (\pi / 2, 0)$ (Hadamard walk).
    (\figlab{a}) Quasienergy spectrum with two energy gaps occurring at energy $\epsilon = 0$ and $\epsilon = \pi$.
    (\figlab{b}, \figlab{c}) The corresponding quasienergy eigenstates of the upper band in the two time frames, Eqs.~\eqref{eq:1dss_time_frame_1}, \eqref{eq:1dss_time_frame_2}, displayed on the Bloch sphere. 
    Chiral symmetry constrains the eigenspinors to lie in a plane, $x=0$, while the quasimomentum is varied across the Brillouin zone performing a closed loop.
	The color gradient indicates the winding direction around the Brillouin zone.
    The (signed) winding number associated with transformation differs in the two time frames, $\nu^{\prime} = 1$ in (\figlab{b}) and $\nu^{\prime\prime} = 0$ in (\figlab{c}).
    The topological invariants of the bulk are given by the sum and difference of the two winding numbers, $(\nu_0, \nu_\pi) = (\nu^{\prime} + \nu^{\prime\prime}, \nu^{\prime} - \nu^{\prime\prime})/2 + 1/2$. See also Fig.~\figref{fig:phase_diagrams}{a} for the related phase diagram.
    }
    \label{fig:winding}
\end{figure*}

\subsection{Topological phases and symmetries}\label{sec:symmetries}

In the context of Floquet theory, the evolution of the quantum state can be expressed by the action of a time-independent effective Hamiltonian $\op{H}$, defined by $\op{W} = \e^{-\imag\hspace{0.3pt}\op{H}}$ \cite{Grifoni1998, Goldman:2014}.
Due to the discrete spatial translational invariance implied by the lattice, the corresponding eigenstates are Bloch waves characterized by a quasimomentum $\vect{k}$, which takes values within the Brillouin zone $(-\pi/a, \pi/a]^N$.
Likewise, the discreteness of the time evolution implies that the eigenvalues of the the effective Hamiltonian $\op{H}$ are quasienergies, denoted by $\epsilon$, which in our notation take dimensionless values in the interval $(-\pi, \pi]$.
Note that physical energy units can be restored trough multiplication by the quantity $\hbar/T$.
In DTQWs, the quasienergy spectrum reveals a band structure with two bands resulting from the two internal states, as can be seen in Fig.~\figref{fig:winding}{a}, where we provide the quasienergy spectrum for the 1D split-step protocol with $(\theta_1, \theta_2) = (\pi / 2, 0)$ (\emph{Hadamard walk}).
For a generic choice of the coin parameters, these two bands are gapped.
The gapped spectrum relates quantum walks to static systems like insulator materials. 
However, unlike in static systems, the Floquet quasienergy spectrum can also have a gap at $\epsilon = \pi$, since quasienergies  $\epsilon=-\pi$ and $\epsilon=\pi$ are identified.
In addition, artificial electric \cite{Cedzich2013,Genske2013} and magnetic fields \cite{Arnault:2016,Gedik:2015} can lead to a higher number of bands, which can possess nontrivial topological properties as well.

Adapting methods developed for static topological insulators to the effective Hamiltonian $H$, Demler et al.\ \cite{Kitagawa2010Expl} have shown that DTQWs can reproduce all ten classes of nontrivial topological phases in one- and two-dimensions for non-interacting particles \cite{kitaev2009periodic,Ryu2010}.
Topological phases can be assigned to different realizations of the effective Hamiltonian and the corresponding topological invariants occur in the form of winding numbers of the Bloch energy eigenstates \cite{Hasan2010}. 

However, a closer inspection of DTQWs reveals that their so-called Floquet topological phases exhibit an even richer structure, which can only be accessed by analyzing the full time evolution of the walk.
This holds for both 1D and 2D DTQWs \cite{Kitagawa2010TopChar, Asboth2012, Rudner2013}.
For instance, the topological phases of the 1D split-step protocol originate from a special symmetry of the walk protocol, which is called \emph{chiral symmetry}.
A walk operator $W$ exhibits chiral symmetry if a unitary operator $\Gamma$ exists, which transforms it as follows: $\op{\Gamma}\, \op{W}\, \op{\Gamma}^{\dagger} = \op{W}^\dagger$ $\Leftrightarrow$ $\op{\Gamma}\, \op{H}\, \op{\Gamma}^{\dagger} = -\op{H}$.
Although the split-step walk operator $\op{W}_\text{1D}$ defined in Eq.~\eqref{eq:1d_protocol} does not have chiral symmetry, one can show that the two walk operators
\begin{align}
    \op{W}_\text{1D}^{\prime} &= \op{C}(\theta_1 / 2)\, \op{S}_x^\downarrow\, \op{C}(\theta_2)\, \op{S}_x^\uparrow\, \op{C}(\theta_1 / 2) 
    \label{eq:1dss_time_frame_1}\,, \\
    \op{W}_\text{1D}^{\prime\prime} &= \op{C}(\theta_2 / 2)\, \op{S}_x^\uparrow\, \op{C}(\theta_1)\, \op{S}_x^\downarrow\, \op{C}(\theta_2 / 2)
    \label{eq:1dss_time_frame_2}\,,
\end{align} 
obtained through a cyclic permutation of the single walk operations, do exhibit chiral symmetry, with the symmetry operator being $\op{\Gamma} = \op{\sigma}_1$ \cite{Asboth2013}.
The cyclic permutation has split the coin operations into two parts, $\op{C}(\theta_i) = \op{C}(\theta_i / 2)\, \op{C}(\theta_i / 2)$, $i = 1,2$.
Since the walk operations repeat themselves periodically, a cyclic permutation of these operations corresponds to a change of basis preserving the underlying topological structure.
Likewise, cyclic permutations allowed identifying \emph{time-reversal symmetry} in Floquet topological insulators \cite{Carpentier2015}.
Hence, the two walk operators in Eqs.~(\ref{eq:1dss_time_frame_1}), (\ref{eq:1dss_time_frame_2}) are chiral-symmetric representations of the same walk, but expressed in two different time frames.
It results from chiral symmetry that each eigenstate at quasienergy $\epsilon$ has a chiral-symmetric partner eigenstate at quasienergy $-\epsilon$.
In particular, if eigenstates exist with quasienergy either $\epsilon=0$ or $\epsilon=\pi$, these states can be their own symmetry partners, i.e., be eigenstates of the symmetry operator $\Gamma$.
This characteristic ensures the robustness of TP edge states in the 1D split-step walk (see Sec.~\ref{sec:TPES}).

We obtain a geometrical representation of the topological twist of the 1D split-step walk by displaying on the Bloch sphere the eigenspinors of the two chiral-symmetric walk operators defined in Eqs.~\eqref{eq:1dss_time_frame_1}, \eqref{eq:1dss_time_frame_2}.
The eigenspinors $\pm \,\vect{n}(k)$ with quasimomentum $k$ are determined by the translational invariant effective Hamiltonian, \raisebox{0pt}[0pt][0pt]{$H=\sum_{k}\epsilon(k)\ket{k}\!\bra{k}\otimes\vect{n}(k)\cdot\vect{\sigma}$}.
It directly follows from chiral symmetry that the eigenspinors with quasienergy $\epsilon\neq 0,\pi$ lie in the plane $x=0$.
This holds true, in particular, for the bulk eigenstates, whose quasienergies lie outside of the gaps, as shown in Fig.~\figref{fig:winding}{a}.
Hence, if we vary the quasimomentum $k$ across the the whole Brillouin zone, the eigenspinor rotates in the plane performing a closed trajectory, winding a (signed) number of times around the origin, as shown in Fig.~\figref{fig:winding}{b,c}.
The difference and sum of the signed winding numbers associated with the two time frames yield a pair $\mathbb{Z} \times \mathbb{Z}$ of topological invariants \cite{Asbth2014, Asboth2012, Jiang2011}.
For the derivation of the winding numbers, the reader is referred to Ref.~\citenum{Asboth2013}.

\begin{figure}
    \includegraphics[width=1\columnwidth]{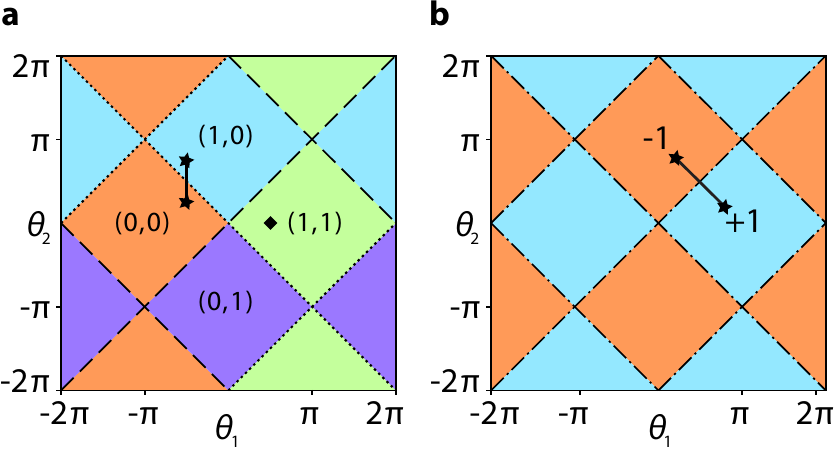}
    \caption{
    Topological invariants assigned to the coin angles of the 1D split-step walk (\figlab{a}) and the 2D protocol (\figlab{b}). 
    Due to the form of the coin operator, $\op{C}(\theta)$, the walk possesses a $4\pi$-periodicity in the coin angles. 
    At the phase boundaries, the gap closes at quasienergy $\epsilon = 0$ (dotted), $\epsilon=\pi$ (dashed), or both at $\epsilon=0$ and $\epsilon=\pi$ (dash-dotted). 
    The coin angle pairs chosen in the numerical examples in this work, and the corresponding phase transitions defined in Eqs.~\eqref{eq:coin_angles_1D}, \eqref{eq:coin_angles_2D} are also displayed (\starSym\!\textbf{---}\!\starSym). 
    The 1D Hadamard walk $(\theta_1, \theta_2) = (\pi/2, 0)$, which is discussed in Fig.~\ref{fig:winding}, is also shown (\diamondSym).\vspace{-2mm}
    }
    \label{fig:phase_diagrams}
\end{figure}

These invariants classify the topological phases of the split-step walk, and depend only on the coin angles $(\theta_1,\theta_2)$, as shown by the phase diagram in Fig.~\figref{fig:phase_diagrams}{a}.
In essence, the pair of topological invariants, $(\nu_0,\nu_\pi)$, count the minimal number of times the band gap closes at quasienergy $\epsilon=0$ and $\epsilon=\pi$, respectively, as the walk is continuously transformed into the topological phase characterized  by $(0,0)$.
Note, however, that the topological protection of these states holds only for perturbations that can be continuously contracted to unity.
For noncontinuous perturbations, instead, the topological phase diagram relies on a single signed winding number, as recently demonstrated in Ref.~\cite{Cedzich:2016}.

In two dimensions, a Floquet topological invariant  $\mathbb{Z}$, the so-called Rudner winding number \cite{Rudner2013}, identifies the topological phases of the 2D DTQW protocol \cite{Edge2015}.
The topological phase diagram is shown in Fig.~\figref{fig:phase_diagrams}{b} as a function of the coin angles.
Remarkably, due to the Floquet character of the DTQW protocol, nontrivial topological phases exist even if the topological invariants assigned to the effective Hamiltonian (i.e., the Chern numbers) are zero.
Moreover, we note that, unlike in one dimension, the 2D DTQW protocol possesses nontrivial topological phases without need for specific symmetries.

\begin{figure*}
    \includegraphics[width=1\textwidth]{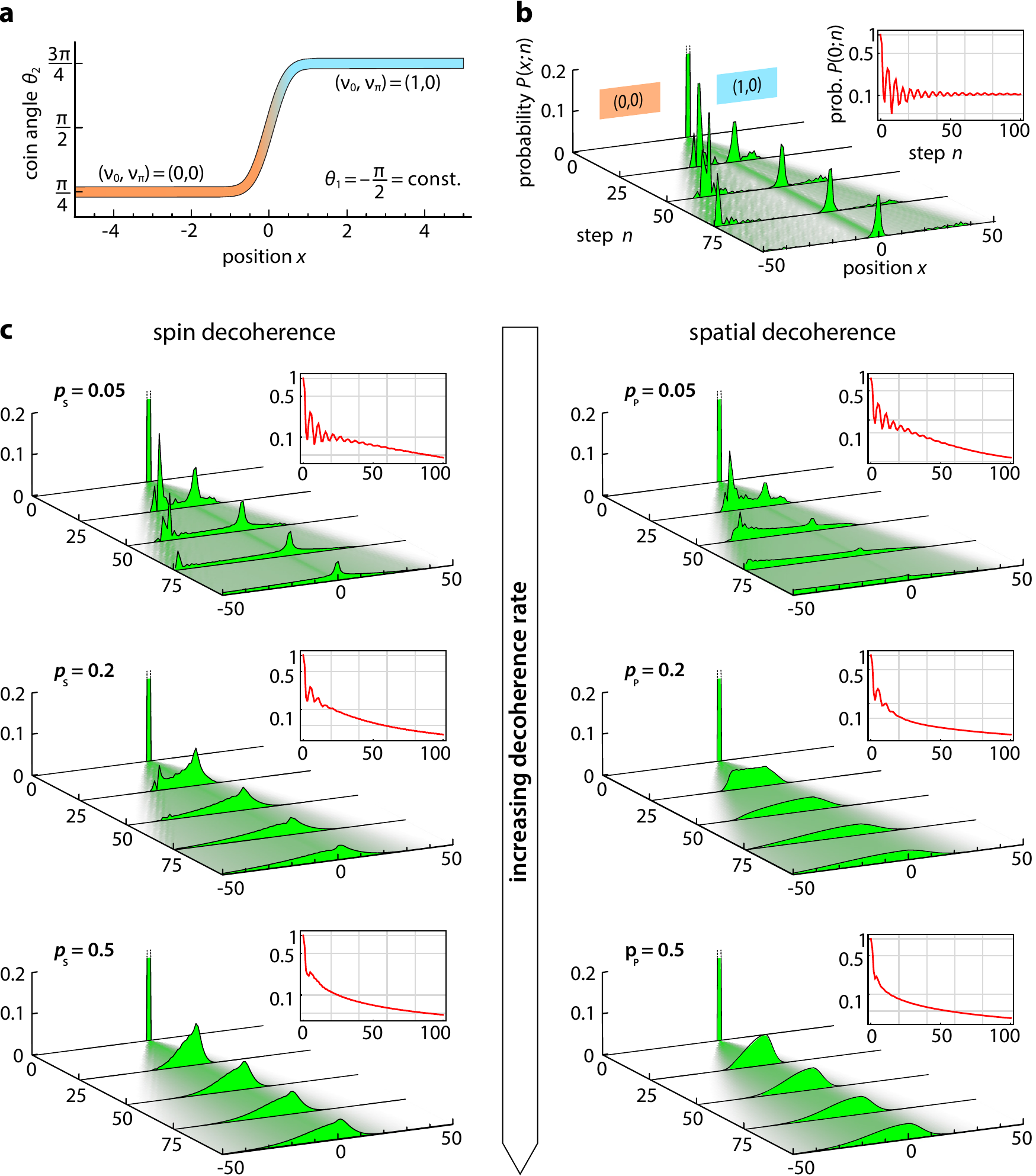}
    \caption{
    (\figlab{a}) 
    Position dependency of the coin angle in the 1D split-step walk given by Eq.~\eqref{eq:coin_angles_1D} realizing two spatially adjacent, distinct topological domains with invariants $(\nu_0, \nu_\pi) = (0,0)$ for $x\ll 0$ and $(1,0)$ for $x\gg 0$. 
    We use a smooth crossover transition corresponding to the diffraction-limited optical resolution of our imaging system (see Sec.\,\ref{sec:experiment} for details).
    (\figlab{b}) 
    Decoherence-free evolution of the spatial density distribution $P({x};n)$ as a function of the number of steps $n$ for a walker initially prepared in the single site state $\ket{0,\downarrow}$. 
    The narrow peak located at the boundary near $x=0$ indicates the component of the walker populating the TP edge state.
    (\figlab{c}) The same walk is subject to pure spin decoherence and pure spatial decoherence with increasing decoherence probabilities $p_\text{S}$, $p_\text{P}$. 
    Insets: time dependence of the walker's probability $P({x=0};n)$ to be at the origin $x=0$ in logarithmic scale. It exhibits an exponential decay for small amounts of decoherence, while stays constant for the decoherence-free evolution.
	The time evolution is calculated for a large number of lattice sites (201) to prevent the walker from reaching the boundaries in the given maximum number of steps.
    }
    \label{fig:decoherence_evolution_1d}
\end{figure*}

\subsection{Topologically protected edge states}\label{sec:TPES}

We consider a spatially inhomogeneous DTQW in which the coin angles depend on the position.
The coin angles are allowed to assume any value inside a spatially confined region at the interface between bulk regions, where the coin angles are kept constant, instead.
When these bulk regions are associated with different topological invariants, TP edge states occur at energies lying in the gaps of the bulk insulators.
More precisely, the bulk-boundary correspondence principle states that the minimum number of edge states is equal to the algebraic difference (in absolute value) between the topological invariants of the individual bulk phases.

For the investigation of TP edge states in the 1D protocol, we choose
\begin{equation}
    (\theta_1,\, \theta_2)= \begin{cases}
                                (-\pi/2,\,\;\, \pi/4)   &  x\ll 0 \\
                                (-\pi/2,\, 3\pi/4)      &  x\gg 0
                            \end{cases}
    \label{eq:coin_angles_1D}
\end{equation}
realizing two spatially adjacent topological phases with invariants $(\nu_0, \nu_\pi) = (0,0)$ for $x\ll 0$ and $(1,0)$ for $x\gg 0$, as delineated in \figref{fig:phase_diagrams}{a}. 
We thus expect a TP edge state with quasienergy $\epsilon=0$ to be localized at the boundary around the site $x=0$. 
To account for realistic experimental conditions, we considered a regular variation of the coin angles over $\simeq\num{2}$ lattice sites, as displayed in Fig.~\figref{fig:decoherence_evolution_1d}{a}, without abrupt changes.
The width of the transition is related to the optical resolution of our experiment, introduced in Sec.\,\ref{sec:experiment}.
Under these conditions, we studied the time evolution of a walker initially prepared in the single-site state 
$\ket{\psi_0} = \ket{0, \downarrow}$.
The results for the ideal situation without decoherence are presented in Fig.~\figref{fig:decoherence_evolution_1d}{b}, where the spatial probability distribution is shown as a function of position $x$ and number of steps $n$, $P(\vect{x};n) = \sum_{s\in \{\uparrow, \downarrow\}}  \left| \braket{\vect{x},s|\psi_n} \right|^2$.
Because the initial state has a large overlap with the TP edge state ($\simeq\num{0.3}$ for the example shown in Fig.~\ref{fig:decoherence_evolution_1d}), the walker is trapped at the boundary with a high probability, yielding a peaked position distribution around the origin even in the long time limit.

\begin{figure}
    \includegraphics[width=0.9\columnwidth]{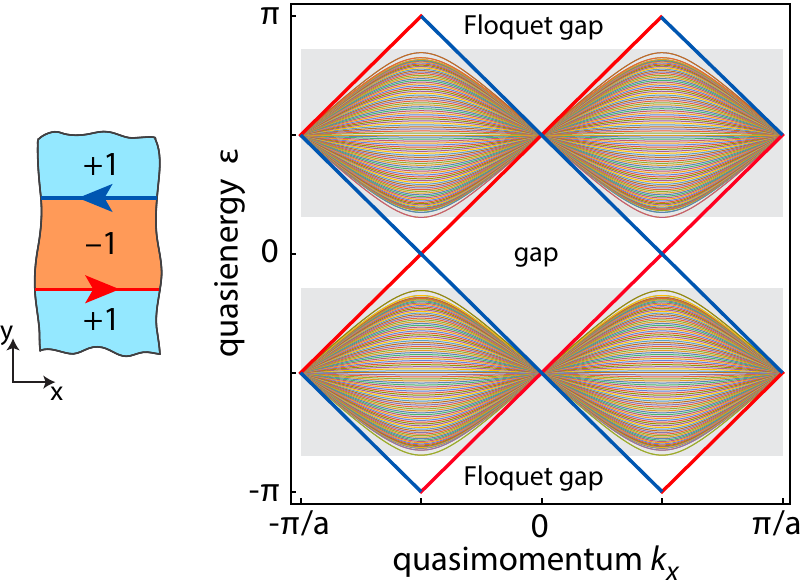}
    \caption{
    Quasienergy spectrum of an inhomogeneous 2D DTQW with a horizontal strip geometry.
	The horizontal strip, 40 sites wide along the $y$-direction, is associated with Rudner invariant $-1$, whereas the rest of the bulk with $+{1}$ (refer to Fig.~\figref{fig:phase_diagrams}{b} for the phase diagram).
    Unidirectional edge modes are visible in the gaps (thick lines), with the blue and red color denoting each edge of the strip.
	For any given quasienergy $\epsilon$ in the gaps, two TP edge modes exist per edge, as expected from the bulk-boundary correspondence principle.
	The spectrum is computed numerically using 100 sites in the $y$-direction.
    }
    \label{fig:2d_edge_spectrum}
\end{figure}

In the 2D walk protocol, the boundary between two distinct topological domains describes a 1D contour.
Along this boundary, which can have in general any shape, TP edge states are expected to exist \cite{Asboth2015}.
However, unlike in the 1D split-step walk, the wavefunction of the TP edge states is delocalized in space, extending along the whole length of the boundary.
As a result of that, a walker in a superposition of TP edge states is no longer confined in the vicinity of the initial site, but can propagate along the whole boundary.
We gather further insight into the transport dynamics along edges by studying the propagation of a wave packet along a straight boundary, which we assume oriented along, say, the $x$-direction.
The flatness of the boundary ensures that the quasimomentum in the boundary's direction, $k_x$, is preserved, so that it can be used to derive the energy dispersion relation of the edge modes.
Fig.~\ref{fig:2d_edge_spectrum} shows the quasienergies as a function of the quasimomentum $k_x$ computed from the effective Hamiltonian for the case of horizontal boundaries between topological domains.
The quasienergy spectrum shows edge modes present in the gaps of the bulk phases.
Recalling the expression of the group velocity, $v_\text{g}(k)=\partial \epsilon(k_x)/\partial k_x$, characterizing the motion of a wave packet, we realize from the the slope of the dispersion relations that the TP edge modes transport currents in a unidirectional manner.
Moreover, for the specific situation of a straight horizontal boundary as considered in Fig.~\ref{fig:2d_edge_spectrum}, it appears that the group velocity does not depend on $k$ (i.e., dispersionless transport), being equal to $\pm\num{1}$ site per step.
We remark that dispersionless transport is not a topological feature, but rather a quantum transport property of the specific DTQW protocol defined in Eq.~\eqref{eq:2d_protocol}.

\begin{figure*}
    \includegraphics[width=\textwidth]{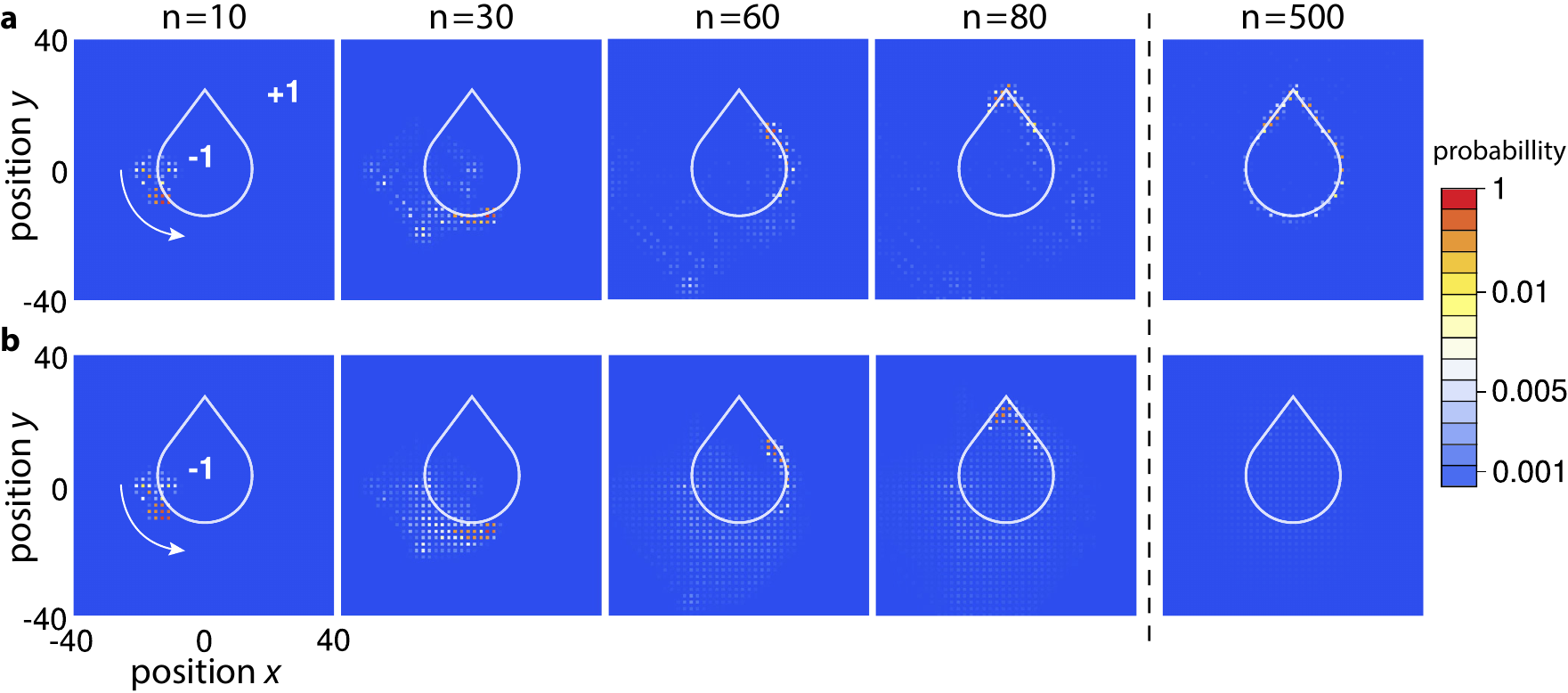}
    \caption{
    (\figlab{a}) Color-coded spatial probability distribution $P(\vect{x};n)$ of a decoherence-free two-dimensional DTQW.
    The coin angles depend on the position as specified by Eq.~\eqref{eq:coin_angles_2D}, creating a droplet-shaped topological island with Rudner invariants $-1$ inside and $+1$ outside of the island.
    The width of the transition is limited by the optical resolution of our experimental setup with Abbe radius $R_\text{A} \simeq 0.8\,a$ (see Sec.\,\ref{sec:experiment} for details).
    The walker is initially prepared in the single site state $\ket{(x=-15,\,y=0),\,\downarrow}$ near the phase boundary and shows a unidirectional moving population of edge states around the boundary as time evolves.
    In (\figlab{b}) the same walk is subject to spin decoherence under realistic experimental conditions ($p_\text{S}=0.05$), exhibiting a slow decay of the edge current over time.
    An animation showing the evolution over $1000$ steps is provided online \cite{droplet_shaped_edge_state_arxiv}.\vspace{-1mm}
	\vspace{-2mm}
    }
    \label{fig:2D_island}
\end{figure*}

To give evidence of the robustness of TP edge modes against deformations of the boundary's shape, we have chosen the boundary to form a closed topological island with a droplet shape, with the coin angles being defined as
\begin{equation}
    (\theta_1,\, \theta_2)= \begin{cases}
                                (\pi/5,\, 4\pi/5)   &  (x,y) \in \text{inside,} \\
                                (4\pi/5,\, \pi/5)   &  (x,y) \in \text{outside.}
                            \end{cases}
    \label{eq:coin_angles_2D}
\end{equation}
With reference to the phase diagram in Fig.~\figref{fig:phase_diagrams}{b}, this choice of angles is associated with Rudner invariants $-1$ inside and $+1$ outside.
We have chosen to add a sharp corner on top of the topological island to test the robustness of the TP edge modes against irregularities of the boundary.
As in the 1D case, we again consider a continuous variation of the coin angle at the boundary.
Angles at the crossover between the inside and outside regions are varied along the line marked in the phase diagram in Fig.~\figref{fig:phase_diagrams}{b}.
Fig.~\figref{fig:2D_island}{a} again shows the spatial probability density distribution $P(\vect{x};n)$ as a function of position $\vect{x}$ and number of steps $n$. 
We initialize the walker in a single site near the boundary, so that its state has a significant overlap with the TP edge states, leading to a unidirectional propagation around the island.
In the absence of decoherence effects, we observe that the edge current persists even after many revolutions around the island, indicating the presence of metallic edge states delocalized along the whole contour of the island.
However, unlike for the straight boundary discussed in Fig.~\ref{fig:2d_edge_spectrum}, which exhibits dispersionless transport, we observe for the droplet-shaped island that the wave packet's probability distribution spreads along the entire border after several revolutions.
We attribute the observed dispersion to the short radius of curvature associated with the border.

\section{Decoherence effects on topologically protected edge states}
\label{sec:robustness}

\subsection{Stroboscopic decoherence model}\label{sec:stroboscopic_model}
Quantum superposition states are fragile against decoherence \textemdash that is, disturbances caused by the surrounding environment onto the quantum system.
The effect of decoherence on the quantum evolution can be effectively described as the projection of quantum states onto a particular basis of so-called \emph{pointer states} \cite{Schlosshauer2007}, which are robust against decoherence.
In quantum-walk experiments with neutral atoms, the pointer states are the spin $\ket{s}, s \in \{\uparrow, \downarrow\}$, and the position states $\ket{\vect{x}}, \vect{x} \in \mathbb{Z}^N$ \cite{Alberti2014}.
Assuming a small amount of decoherence per step, we can approximate the continuous-time decoherence process through a series of discrete measurement operations, which are applied stroboscopically after each unitary step of the walk.
We assume that each measurement only resolves the walker's state with a certain {decoherence probability} $0\leq p \leq 1$.
The walk's evolution is coherent for $p=0$, while it describes a classical random walk for $p=1$.
Our model relies on the assumption of small decoherence to be accurate, $p\ll1$.
Henceforth, we denote by $p=p_\text{S}$ and $p=p_\text{P}$ the decoherence probability related to the spin and position states, respectively.

We follow Ref.~\citenum{Alberti2014} to describe the non-unitary time evolution of the walker by means of the reduced density matrix formalism.
As the walker is initially prepared in a pure state $\ket{\psi_{0}}$, the initial density matrix is $\op{\rho}_0 = \ket{\psi_0}\! \bra{\psi_0}$. 
The density matrix $\op{\rho}_{n+1}$ describing the walker at time $t=(n+1)\,T$ depends only on the state of the walker at time $t=n\,T$ (Markovian assumption).
Hence, $\op{\rho}_{n+1}$ is obtained through the repetitive application of the linear superoperator $\mathcal{E}$, which accounts for the effect of environment-induced decoherence at each step \cite{preskill1998lecture}:
\begin{align}
    \op{\rho}_{n+1} &=  \op{\mathcal{E}}^{n+1}(\op{\rho}_{0}) = \op{\mathcal{E}}(\op{\rho}_{n}) =\nonumber \\ &= (1-p)\, \op{W}\,
    \op{\rho}_{n}\, \op{W}^\dagger + p\, \sum_{i} \op{\mathbb{P}}_i\,
    (\op{W}\, \op{\rho}_{n}\,
    \op{W}^\dagger)\, \op{\mathbb{P}}_i^\dagger,
    \label{eq:time_evolution_decoherence}
\end{align}
where $i\in\{\uparrow, \downarrow\}$ for pure spin and $i \in \{\vect{x}\}$ for pure position decoherence. 
The projectors $\op{\mathbb{P}}_i$ are defined as
\begin{equation}
	\label{eq:projectors}
    \op{\mathbb{P}}_{\vect{x}} = \sum_s \ket{\vect{x}, s}\!\bra{\vect{x}, s}\,, 
    \quad
    \op{\mathbb{P}}_{s} = \sum_{\vect{x}} \ket{\vect{x}, s}\!\bra{\vect{x}, s}.
\end{equation}
We found in a previous study that this simple model reproduces in a satisfactory manner the effects of decoherence occurring in our expe\-riments with neutral atoms \cite{Alberti2014}.
In particular, our previous analysis revealed that spin decoherence is the main mechanism responsible for the loss of coherence in the current 1D quantum-walk setup.
We therefore focus in this work primarily on decoherence by spin dephasing.
In addition, our numeric analyses assume a conservative decoherence probability of $p_\text{S} \simeq \num{0.05}$ per step, which is based on previous experimental results \cite{Alberti2014}.
However, the construction of a new quantum-walk setup for 2D DTQWs is underway that promises decoherence probabilities as low as $p_\text{S}<0.01$ owing to a number of technical improvements, including, among others, shielding of stray magnetic fields and suppression of polarization distortions of the optical lattice laser beams.

\vspace{-2mm}
\subsection{Decoherence effects on TP edge states in 1D}\label{sec:decoh_TP_1D}

We illustrate the effect of decoherence by analyzing the walk evolution of a 1D DTQW with two adjacent bulks with coin angles defined by Eq.~\eqref{eq:coin_angles_1D}.
We again initialize the walker in a single site state $\ket{0, \downarrow}$ near the boundary, so that the walker is able to populate the TP edge state.

In Fig.~\figref{fig:decoherence_evolution_1d}{c} we show the spatial probability distribution
$P(\vect{x};n) = \sum_{s\in \{\uparrow, \downarrow\}} \braket{\vect{x},s|\op{\rho}_n|\vect{x},s}$
obtained numerically using Eq.~\eqref{eq:time_evolution_decoherence}.
The resulting distribution of the walk reflects two phenomena.
First, the walker occupies the TP edge state, resulting in a narrow probability peak located around the crossover point at $x=0$. 
Second, this peak stays nearly constant in position and shape, but decays over time with a rate  increasing with the decoherence strength, $p$.
On the other hand, the component of the walker's wave function that has no overlap with the TP edge state expands in the bulk.
For small decoherence, the expansion preserves a ballistic-like behavior for many steps, resulting in the characteristic distribution with off-center peaks.
The number of peaks and the direction of propagation depends on the initial state of the walker.
For stronger decoherence, this expansion exhibits a diffusive behavior \cite{Alberti2014}, with a distribution centered around the starting point, thus overlapping with the TP edge state.
From our simulations, it results that experiments must be conducted under small decoherence conditions, $p<0.05$, in order to be able to detect the persistence of a sharply peaked distribution at the boundary \textemdash a signature of the TP edge state.
It should be noted that the decoherence rate determines the point in time where the expansion changes from a ballistic spreading on a short time scale to a diffusive behavior for longer times \cite{Alberti2014}.
 
The probability for the walker to remain in the origin, $P({x=0};n)$, is an indicator for the robustness of the TP edge state, see the insets in Fig.~\ref{fig:decoherence_evolution_1d}.
It shows an oscillatory evolution for a short transient due to the dynamics of the walker's component overlapping with the bulk states, which is free to expand into the bulk.
For longer times, the probability stays constant for the decoherence-free evolution, while decays nearly exponentially for small decoherence rates.
In case of strong decoherence, the population of the TP edge state deviates from a simple exponential decay.
In this regime, however, the assumption underlying our stroboscopic decoherence model, $p\ll 1$, does not hold anymore, see Sec.~\ref{sec:stroboscopic_model}.
A more detailed discussion based on an analytic model is presented in Sec.\,\ref{sec:analytic}.

\vspace{-1mm}
\subsection{Decoherence effects on TP edge states in 2D}
\begin{figure}
    \includegraphics[width=\columnwidth]{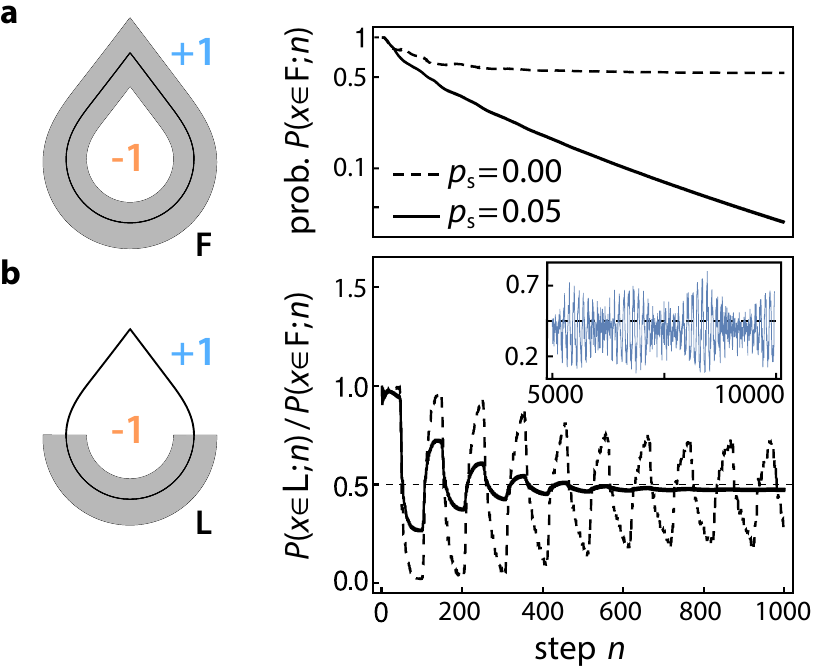}
	\vspace{-5mm}
    \caption{
    (\figlab{a}) Probability $P({\vect{x}\in\mathrm{F}};n)$ for the walker to be inside the grayed region F as a function of the number of steps, $n$, in logarithmic scale. 
    (\figlab{b})~Probability $P({\vect{x}\in\mathrm{L}};n)$ for the walker to be inside the grayed region L near the lower half of the phase boundary, normalized to the population probability $P({\vect{x}\in\mathrm{F}};n)$.
    The probabilities are shown for the unitary walk evolution (dashed curves) and for a decoherence rate $p_\text{S} = \num{0.05}$ (solid curves).
    Inset: close-up view in the long time limit for the evolution without decoherence.
	\vspace{-2.5mm}
    }
    \label{fig:edge_population_probablity_2D}
\end{figure}

The evolution of the 2D walk revolving around the droplet-shaped topological island in the presence of weak spin decoherence is presented in Fig.~\figref{fig:2D_island}{b}. 
The probability current along the boundary shows a slow decay over time.
As an indicator for the population of the TP edge modes, we study the probability $P({\vect{x}\in\mathrm{F}};n)$ for the walker to be situated in a small band, $F$, around the edge, as shown in Fig.~\figref{fig:edge_population_probablity_2D}{a}.
For an initial transient period of $\simeq 50$ steps, the edge probability shows a decrease which is nearly independent of the decoherence probability, and is attributed to the non-vanishing projection of the initial single-site state onto the bulk states.
For the decoherence-free evolution, the probability tends, in the long time limit, to a constant value, $P({\vect{x}\in\mathrm{F}};n\gg1) = \num{0.53}$.
It is worth emphasizing that such a high probability is favorable to future experiments, which aim to detect matter waves trapped at the boundary.
In the presence of decoherence, instead, we observe an approximately exponential decay in qualitative agreement with the results obtained in the 1D walk (see Sec.~\ref{sec:decoh_TP_1D}).

While decoherence reduces the probability current, it has no discernible effect on the propagation velocity of a wave packet along the boundary.
The comparison between Fig.~\figref{fig:2D_island}{a} and \figref{fig:2D_island}{b} shows, in fact, that the front of the wave packet moves, in both cases, with a speed of approximately one lattice site per step, irregardless of whether the walker is subject to decoherence.
This velocity is also in good agreement with that computed in Sec.~\ref{sec:TPES} from the energy dispersion relation of a flat boundary.
Interestingly, the propagation along the boundary attains the highest velocity, one site per step, allowed by the 2D quantum walk protocol defined in Eq.~\ref{eq:2d_protocol} (i.e., attains the effective speed of light for the DTQW protocol).

To gain further insight into the dynamics of the walker revolving around the island, we display in Fig.~\figref{fig:edge_population_probablity_2D}{b} the probability $P({\vect{x}\in\mathrm{L}};n)$ for the walker to be in the lower half, L, of the boundary.
This probability exhibits periodic oscillations in time with a period that is independent of the decoherence rate, and approximately equal, in units of steps, to the length of the contour of the topological island.
The period, in particular, corroborates our previous observation that the wave packet moves unidirectionally along the boundary with a velocity of nearly one site per step.
We also observe that the oscillation amplitude is damped after several revolutions.
We explain this damping as the result of the group velocity dispersion of the TP edge states, which make the wave packet spread along the entire boundary.
In the presence of decoherence, the damping occurs on a much shorter time scale, presumably due to the walker's component that is diffused into the bulk, but yet located inside the band $L$.
For the unitary evolution, however, oscillations persist with the same periodicity for long times, as shown in the inset of Fig.~\figref{fig:edge_population_probablity_2D}{b}.
The modulation of the oscillation amplitude over long time scales is attributed to partial collapses and revivals, since the time evolution is unitary and the edge of the topological island constitutes a finite Hilbert space with a discrete spectrum \cite{Bocchieri:1957}.
A detailed study of the residual oscillations would require further investigation.

\subsection{Analytical model of the decay of TP edge states}
\label{sec:analytic}
We consider the 1D split-step walk protocol to derive a simple analytical model predicting the decay rate of TP edge states in the presence of decoherence.
Assuming that the walker is initially in a TP edge state $\ket{E}$, we compute the probability $\Pi(n)$ that it remains in the same state after $n$ steps.
Due to decoherence, the walker's wave function acquires a non-vanishing overlap with the continuum of the bulk states.
In order to carry out the computation analytically, we assume that the walker's component coupled to the bulk rapidly leaves the boundary because of the nearly ballistic expansion, without ever repopulating the TP edge state.
Under this assumption, which is well justified in the regime of weak decoherence $p\ll1$, we obtain in Appendix~\ref{sec:decay_of_tpes_under_decoherence} that the probability of occupying the edge state is
\begin{equation}
    \Pi(n) = \mathrm{tr}\left( \ket{E}\!\bra{E}\, \op{\rho}_n \right) \simeq
    (1-\gamma)^n ,
    \label{eq:edge_population_probability}
\end{equation}
where the decay rate $\gamma$ depends on $\ket{E}$ and is linear in $p$.
For pure spin decoherence, the decay rate is given by
\begin{equation}
		\gamma_\text{S} = p_\text{S}\, \biggl[ 1 - \sum_s \Bigl(\sum_{\vect{x}}
        \left| \braket{\vect{x},s|E} \right|^2 \Bigr)^2 
        \biggr].
    \label{eq:loss_rate}
\end{equation}
A similar expression for the decay rate $\gamma_\mathrm{P}$ for pure position decoherence is provided in Appendix~\ref{sec:decay_of_tpes_under_decoherence}.
Moreover, the expression in Eq.~\eqref{eq:loss_rate} can be written in a more compact form as $\gamma_\text{S}= p_\text{S}\, ( 1 - \sum_s \left|\braket{s|s_E}\right|^4 )$ by exploiting the factorization of 1D TP edge states into a position and spin component, $\ket{E}=\ket{\chi}\otimes\ket{s_E}$, as ensured by chiral symmetry (see Sec.~\ref{sec:symmetries}).

\begin{figure}
    \includegraphics[width=1\columnwidth]{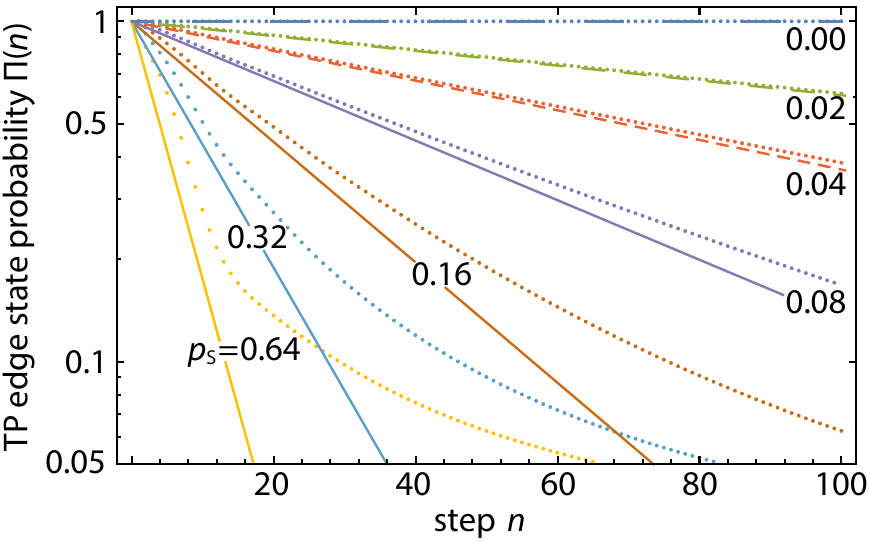}
	\vspace{-3mm} 
    \caption{
	Probability of populating a TP edge state as a function of the number of steps $n$ for different amounts of spin decoherence $p_\text{S}$ (semi-logarithmic scale).
    The data points are calculated numerically for the 1D split-step walk with the coin angles as defined in Eq.~\eqref{eq:coin_angles_1D}, and with the initial state being the TP edge state with quasienergy $\epsilon=0$.
	The solid lines represent an exponential decay as predicted by the analytical model in Eq.~\eqref{eq:edge_population_probability}.
	\vspace{-1mm} 
    }
    \label{fig:edge_population_probability}
\end{figure}

This simple model predicts an exponential decay of the edge state population, which agrees well with the numerical simulations for short times and small decoherence, as shown in  Fig.~\ref{fig:edge_population_probability}.
In addition, we attribute deviations from the exponential decay model, observed for longer times, to a non-negligible probability that decoherence transfers the walker from the bulk states back to the TP edge state.

\subsection{Limits of the stroboscopic decoherence model}
In Sec.~\ref{sec:stroboscopic_model}, we have modeled the effect of decoherence through a single measurement operation, of either the spin or the position of the particle, applied after each coherent step of the walk $\op{W}$.
This constitutes, in general, a good approximation of the actual dynamics, provided that the amount of decoherence is small ($p \ll 1$), as is the case of ultracold atom experiments (see Sec.~\ref{sec:experiment}).

However, situations exist where the stroboscopic application of decoherence can completely fail to describe the decay of a TP edge state.
We would like to caution the reader about that by providing an explicit example, which is constructed \emph{ad hoc} to prove the existence of a TP edge state that is robust against any amount of stroboscopic spin decoherence.
Such a situation can occur when the quantum walk possesses a special symmetry (for example, chiral symmetry) that forces the spin component of the TP edge state to be oriented along a given direction, for example, along the $z$-direction.
It is evident in this case that spin measurements in the $z$-basis leave the TP edge state unperturbed.
This is confirmed by Eq.~\eqref{eq:loss_rate}, predicting in this case a decay rate $\gamma_S=0$ for any $p_\text{S}$.

This can be realized by considering a unitary transformation of the walk operator in Eq.~\eqref{eq:1dss_time_frame_1},  
$\op{\tilde{W}}_\text{1D} = \op{C}(\pi/2)\, \op{W}^\prime_\text{1D}\, \op{C}(-\pi/2)$.
This transformation is equivalent to a cyclic permutation, and it does not change the walk evolution in the bulk as well as the corresponding topological invariants.
The chiral symmetry operator of the transformed walk is $\sigma_z$ since $\op{\sigma}_z\, \op{\tilde{W}}_\text{1D}\, \op{\sigma}_z = \op{\tilde{W}}_\text{1D}\!^{\dagger}$.
Since the TP edge states are eigenstates of the symmetry operator (see Sec.~\ref{sec:symmetries}), their spin must be either $\ket{\uparrow}$ or $\ket{\downarrow}$, and a projective measurements of the spin in the $z$-basis leave the TP edge state unaffected.
We note that an analogous situation can be reproduced in the Su-Schrieffer-Heeger topological model, where it is known that the sublattice symmetry (tantamount to chiral symmetry) forces the TP edge state to lie on either one of the two sublattices \cite{Asboth:Book}.
Hence, a quantum non-demolition measurement of the sublattice would leave, in like manner, the TP edge state unaffected.
 
A remedy to avoid such seemingly paradoxical situations, where TP edge states are left unmodified by environment-induced decoherence, consists in modifying Eq.~\eqref{eq:time_evolution_decoherence} to allow the decoherence Kraus operators to act after each discrete operation of the single step.
Furthermore, identifying the exact operator-sum representation in terms of Kraus operators of the decohered coin operation would ultimately provide the most accurate modeling of decoherence effects \cite{Andersson:2007}.

\section{Experimental Proposal with Neutral Atoms in Optical Lattices}
\label{sec:experiment}

\subsection{Optical lattice experimental setup}
\begin{figure}
    \includegraphics[width=\columnwidth]{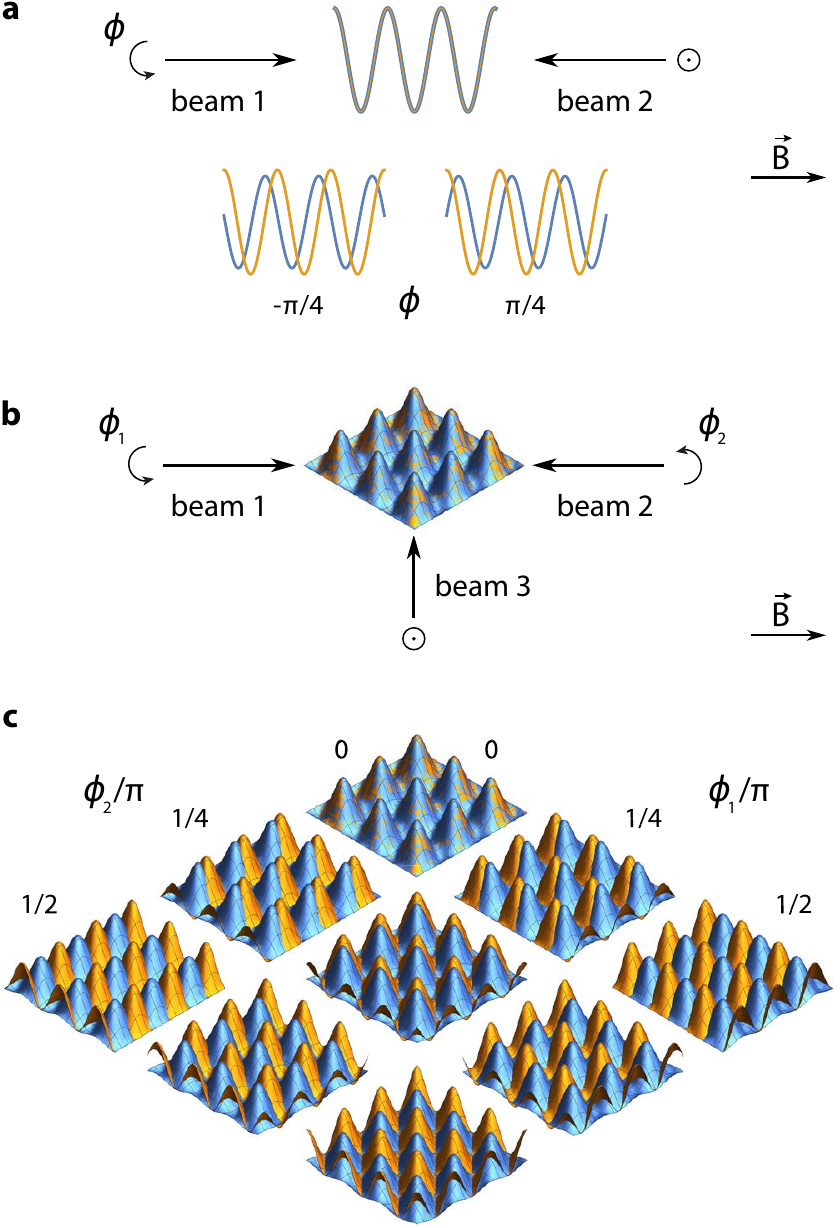}
    \caption{
    (\figlab{a}) 
	 One-dimensional lattice potentials created by two linearly polarized beams.
    A polarization rotation by $\phi$ leads to a relative displacement of the two optical potentials (orange and blue curves), which spin-dependently trap atoms in either the $\ket{\uparrow}$ or $\ket{\downarrow}$ internal state.
	The vector $\vect{B}$ represents the direction of the external magnetic field, which fixes the quantization axis.
    (\figlab{b}) 
    Two-dimensional lattice potentials created by three interfering laser beams for spin-dependent transport on a square lattice.
   	The polarization of beam 3 points out of the plane, whereas the polarization of beam 1 and 2 can rotate, producing spin-dependent displacements along two diagonal directions at $\pm\SI{45}{\degree}$ relative to the quantization axis.
	Two counter-propagating beams (not shown) orthogonal to the plane provide the confinement in the third direction.
    (\figlab{c}) 
    Potential depth of the two spin-dependent optical lattices (orange and blue) for different polarization angles, $\phi_1$  and $\phi_2$.
	}
    \label{fig:transport}
\end{figure}

We have shown in previous experiments \cite{Karski2009} that an atomic quantum walk can be realized employing a single neutral cesium atom in an optical lattice at a specific wavelength $\lambda_{L} = \SI{866}{\nano\meter}$.
The outermost hyperfine ground states,  $\ket{\uparrow} = \left| F=4, m_F=4 \right\rangle$ and $\ket{\downarrow} = \left| F=3, m_F=3 \right\rangle$, define the pseudo spin-1/2 states of the quantum walker.
Due to their different ac-polarizability, each of these states experiences, to a large extent, only the trapping potential of either one of two distinct $\sigma^+$- and $\sigma^-$-circularly polarized optical lattices.
The setup for spin-dependent shift operations in one dimension is depicted in Fig.~\figref{fig:transport}{a}, where two counter-propagating laser beams of linear polarization form a 1D optical lattice along the direction of the quantization axis.
Spin-dependent shift operations are then realized by controlling the polarization and phase of just one of the two optical lattice beams (beam 1 in the figure).
A rotation of its linear polarization, which is achieved through a shift of the relative phase between circular polarization components, displaces into opposite directions the two circularly polarized optical lattices and, thereby, atoms in different internal state.
Previous implementations \cite{Mandel2003b,Belmechri2013} of this concept based on an electro-optic device suffer from the shortcoming that shift operations are limited to a maximum distance of about one lattice site at a time and, most importantly, to only relative displacements between $\ket{\uparrow}$ and $\ket{\downarrow}$ spin components.
Sole relative displacements are not sufficient to realize the $\op{S}_x^\downarrow$ and $\op{S}_x^\uparrow$ operations, which are required by the split-step walk protocol in Eq.~\eqref{eq:1d_protocol}.
However, we recently demonstrated a different technique for precision polarization synthesis, which overlaps two fully independent laser beams with opposite polarizations to form a beam of arbitrary polarization and phase \cite{Robens2016}.
The new implementation of spin-dependent transport allows us to independently shift each individual spin component by an arbitrary distance, ultimately limited by the Rayleigh length.

We propose to extend the concept of spin-dependent transport, which has hitherto been demonstrated only in one dimension, to a square lattice in two dimensions.
We employ three interfering laser beams with linear polarization, as illustrated in Fig.~\figref{fig:transport}{b}.
With reference to the figure, the polarization of beam 1 and 2 can be rotated in time by an angle $\phi_1$ and $\phi_2$, respectively, employing our recently developed polarization-synthesis setup for each of the two beams.
The polarization of beam 3 is instead fixed and orthogonal to the quantization axis, which is chosen along the direction of beam 1 and 2.
In essence, a rotation of the two polarization angles results in a spin-dependent shift operation along one of the two diagonal directions, as shown in Fig.~\figref{fig:transport}{c}.
This novel experimental scheme allows the precise control of discrete-time spin-dependent shift operations along the two main directions of a square lattice.
We note that our scheme differs substantially from other experimental schemes for continuous-time spin-orbit coupling, which are based on either a dynamical rotation of the magnetic field (i.e., of the quantization axis) \cite{Luehmann:2014} or a dynamical modulation of a magnetic field gradient \cite{Struck:2014,Jotzu:2015}.

The geometric arrangement of laser beams in Fig.~\figref{fig:transport}{b} increases the spacing between adjacent lattice sites by a factor $\sqrt{2}$ (thus, $a = \sqrt{2}\,\lambda_{L}/2$) compared to the 1D lattice presented in Fig.~\figref{fig:transport}{a}, constituting an advantage to optically address each lattice site individually.
In addition, the concurrent interference of the all three beams yields a trap depth that is $3/2$ times as deep as that obtained by a 1D lattice for the same optical power.

\begin{figure}
    \includegraphics[width=\columnwidth]{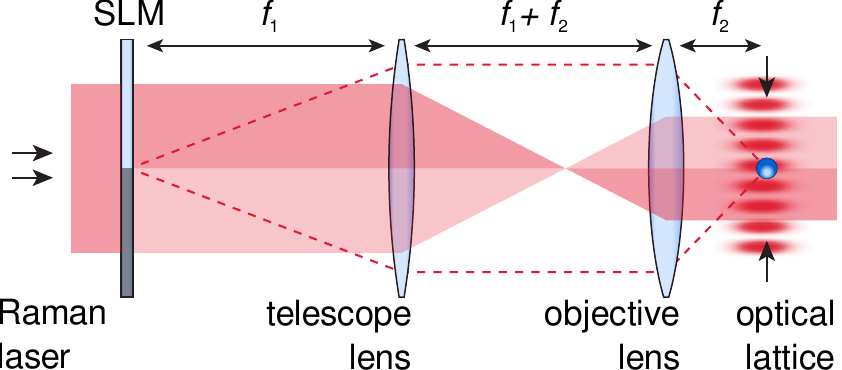}
    \caption{
	The intensity of Raman lasers, utilized to implement the coin operation, is modulated in space  to give rise to sharp topological phase boundaries.
	A spatial light modulator (SLM) creates a structured intensity pattern, which is imaged onto the optical lattice by a high-numerical-aperture (NA=0.92) objective lens mounted in a 4$f$ optical system.
    }
    \label{fig:slm}
\end{figure}

The construction of the experimental apparatus is currently underway.
An objective lens with large numerical aperture (NA), which is placed at $\SI{150}{\micro\meter}$ in front of the 2D lattice, allows us to detect the location of atoms with single site resolution by fluorescence imaging on the D2 line at $\lambda_\text{f}=\SI{852}{\nano\meter}$  \cite{Alberti:2016}, as well as to project a structured intensity pattern for local, optical control of the coin operation.
The coin operation can be implemented either through microwave radiation resonant with the hyperfine splitting at $\SI{9.2}{\giga\hertz}$, or through a pair of Raman laser beams with wavelength $\lambda_\text{C} = \SI{894}{\nano\meter}$ slightly detuned from the D1 line.
Microwave pulses are most suited for driving coin operations with position-independent coin angles, while Raman laser pulses allow spatial variations of the coin angles by modulating their intensity.
For the local control of the Raman laser intensity with single site resolution, we propose the 4f optical system illustrated in Fig.~\ref{fig:slm}.
The coin rotation angle at a certain lattice site depends linearly on the intensity of Raman lasers illuminating that given site.

\subsection{Realization of topological phase boundaries}
\begin{figure}[b]
    \includegraphics[width=\columnwidth]{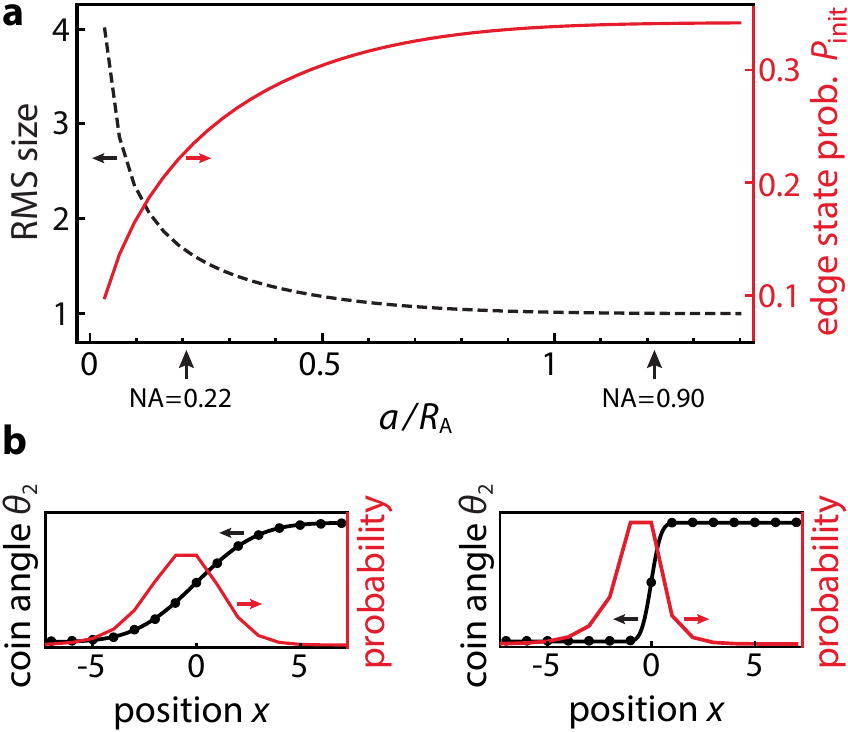}
    \caption{
    Analysis of a TP edge state $\ket{E}$ in the 1D split-step DTQW with coin angles given by Eq.~\eqref{eq:coin_angles_1D} for different slopes of the phase crossover, as determined by the diffraction parameter $a / R_\text{A}$. 
    (\figlab{a}) RMS size of the TP edge state (black, dashed) and overlap probability of the initial state $\ket{x=0,s_E}$ with the TP edge state $\ket{E}=\ket{\chi}\otimes\ket{s_E}$ (red, solid).
	The two vertical arrows indicate the values corresponding to the 1D and 2D quantum-walk setups.
    (\figlab{b}) Coin angles $\theta_2$ (black circles) and position distribution $\sum_s |\braket{E|x, s}|^2$ of the TP edge state (red lines) computed for the current 1D (left) and the new 2D experimental setups (right).
    }
    \label{fig:size_and_population_prob}
\end{figure}

In the experiments, sharp crossovers between topological phases are preferable because their TP edge states are strongly localized in the proximity of the boundary, thereby avoiding slowly decaying tails in the direction of the bulk.
This ensures a relatively high probability that an atom originally prepared in a single lattice site next to the boundary populates the edge state.
Additionally, sharp boundaries make it less demanding for experiments to realize coherence lengths \footnote{The coherence length of DTQWs depends, in general, on the decoherence rate and, for small spin decoherence, is inversely proportional to $p_C$  \cite{Alberti2014}.} longer than the size of TP edge states.

However, there is a limit on how sharp crossovers between different topological domains can be, which is determined by diffraction in the optical system.
For diffraction-limited optical systems, the sharpness of the phase crossover depends on the numerical aperture $\mathrm{NA}$ of the objective lens, the lattice constant $a$, and the wavelength $\lambda_\text{C}$ of the Raman lasers.
Mathematically, the intensity profile experienced by atoms results from the convolution of the profile generated by the spatial light modulator (see Fig.~\ref{fig:slm}) with the point spread function (PSF) of the imaging system \cite{Alberti:2016}.
In the numerical simulations presented in this work, we approximated the experimentally measured Airy-disc-like PSF with a Gaussian function with standard deviation $(\sqrt{2} / \pi) R_\text{A}$, where $R_\text{A} = \lambda_\text{C} / (2\, \mathrm{NA})$ is the Abbe radius.
Hence, the unit step profile with coin angles $\theta_\text{L}$ for $x\leq 0$ and $\theta_\text{R}$ for $x> 0$, which we considered for the 1D simulations, results after the convolution in
\begin{equation}
    \theta(x) = \theta_\text{L} + \frac{\theta_\text{R} - \theta_\text{L}}{2} \left[1 + \mathrm{erf}\left(\frac{a\,\pi}{2\, R_\text{A}}\, x\right)\right],
    \label{eq:coin_angle_function_1d}  
\end{equation}
where $\mathrm{erf}$ is the Gaussian error function.
The present 1D quantum-walk setup with $\mathrm{NA}= \num{0.22}$ \cite{Alberti:2016} and $a = \lambda_{L} / 2$ allows only moderately sharp boundaries, $ R_\text{A} \simeq  4.8 \,a$.
The new 2D quantum-walk setup, instead, features an objective lens with a higher numerical aperture, $\mathrm{NA}= \num{0.92}$, and a longer lattice constant, $a =\sqrt{2}\, \lambda_{L}/2$, resulting in $R_\text{A} \simeq 0.8\,a$.
This permits nearly abrupt phase boundaries, where the coin angle is varied across just $\simeq{1}$ lattice site.

In order to obtain a quantitative relation between the optical resolution of the optical system and the shape of TP edge states, we numerically studied the phase crossover in the 1D protocol as a function of the ratio $a/R_\text{A}$.
As shown in Fig.~\ref{fig:size_and_population_prob}, the size of the TP edge state decreases monotonically with the optical resolution until it attains a constant value around one lattice site.
The figure also displays the probability $P_\text{init} = \left|\braket{E|{x}_0, s_0}\right|^2$ to populate the TP edge state $\ket{E}$ from the initial state $\ket{{x}_0, s_0}$.
In the experiments, it is important to maximize this probability by choosing a sharp boundary and 
the initial spin, $\ket{s_0}$, such that it coincides with the spin of the edge state at position ${x}_0$.
The initial spin can be easily prepared by applying a suitable microwave pulse.

\section{Outlook and discussion}

In this paper, we have studied the robustness of TP against environment-induced decoherence, which causes dephasing of the quantum-walk states.
We have analyzed the effect of decoherence on the existence and form of TP edge states. 
We have found that decoherence of spin and position states leads, in both cases, to an approximately exponential decay of the TP edge state into the bulk states.
A study of phase coherence properties of matter waves propagating along a quantum circuit of TP edge states will be the subject of future work, similar to that pursued by Ref.~\citenum{Levkivskyi:2012} with IQHE solid-state devices \cite{Ji:2003}.

The novel scheme for 2D spin-dependent transport combined with Raman laser pulses to drive the coin operation will allow us to realize arbitrary topological domains in 1D and 2D quantum walks under realistic decoherence conditions.
Owing to a high numerical aperture, the diffraction-limited optical system utilized to project the Raman pulses reduces the size of the TP edge states to a minimum, yielding a high probability to populate them from a single site.

Exploring the limits of the stroboscopic decoherence model revealed that specific TP edge states can be unaffected by decoherence.
In the future, we plan to build upon this result to construct Kraus operators that can pump the walker into a TP edge state when applied periodically in time.
This would allow us to engineer dissipation to protect TP edge states not only from static disorder, but also from a weak amount of environmental decoherence \cite{Bardyn2013}.

As yet, only little is known about the role of interactions in topological insulators \cite{Chen:2012,Grusdt:2013}.
While topological phases of non-interacting systems are relatively well understood, the classification of interacting topological phases is in its infancy.
The most promising direction of future quantum-walk experiments with neutral atoms consists in exploiting the strong, controllable interactions between atoms in order to understand topological phases with interacting particles.
Atoms have in fact the potential to shed new light on topological phases with strongly correlated particles, which go beyond a purely wave-mechanical picture as that of non-interacting topological phases \cite{Wang:2009,Kane2013,Susstrunk:2015}.

\vspace*{2mm}

\begin{acknowledgments}
We thank Carsten Robens, Geol Moon, and Michael Fleischhauer for insightful discussions.
We acknowledge financial support from the Deutsche Forschungsgemeinschaft SFB project Oscar, the ERC grant DQSIM, and the EU project SIQS.
We acknowledge support by the Hungarian Scientific Research Fund (OTKA) under Contract No.\ NN109651, the Deutscher Akademischer Austauschdienst (TempusDAAD Project No.\ 65049).
T.G.\ was supported by the Studienstiftung des deutschen Volkes.
J.K.A.\ was supported by the Janos Bolyai Scholarship of the Hungarian Academy of Sciences.

\end{acknowledgments}

\ifusebibfile
	\bibliographystyle{apsrev4-1}
    \bibliography{topology-decoherence}
\else
\fi
\begin{appendix}

\section{Analytical decay model of TP edge state under decoherence, \texorpdfstring{Eqs.~\hyperref[eq:edge_population_probability]{(\ref{eq:edge_population_probability})} and \hyperref[eq:edge_population_probability]{(\ref{eq:loss_rate})}}{}}

\label{sec:decay_of_tpes_under_decoherence}

We derive an analytical model describing the decay of the TP edge state under pure spin decoherence. 
A model describing the decay under decoherence affecting the position states only, can be derived analogously.

Let $\ket{E}$ be a TP eigenstate of the walk operator $\op{W}$ with quasienergy $\epsilon$. 
The corresponding density matrix $\op{\rho}_0 = \ket{E}\!\bra{E}$ is then invariant under application of the walk operator $\op{W}$:
\begin{equation}
    \op{W}\, \ket{E}\!\bra{E}\, \op{W}^\dagger 
    = \e^{-\imag\,\epsilon}\, \ket{E}\!\bra{E}\, \e^{\imag\,\epsilon}
    = \op{\rho}_0 .
\end{equation}
We consider the 1D walk evolution of this state under spin decoherence as defined by Eq.~\eqref{eq:time_evolution_decoherence}. 
After one step, the walker's state is described by
\begin{equation}
    \op{\rho}_1
        = (1-p_\text{S})\, \op{\rho}_0\, + p_\text{S}\!\! 
        \sum_{s \in \{\uparrow, \downarrow\}} 
        \op{\mathbb{P}}_{s}\, \op{\rho}_0\,\op{\mathbb{P}}_s^\dagger \,,
\end{equation} 
where $\op{\mathbb{P}}_s$ is the projector onto the spin state $s$, as defined in Eq.~\eqref{eq:projectors}.
The probability $\Pi(1)$ to find the walker in the same state $\ket{E}$ is given by
\begin{align}
    \Pi(1) 
    &= \mathrm{tr}\left( \ket{E}\!\bra{E}\, \op{\rho}_1\right) 
        \nonumber\\
    &= (1-p_\text{S})\,\mathrm{tr}\left({\op{\rho}_0}{}^2\right) + p_\text{S}\, 
        \sum_s 
        \mathrm{tr}\left( \op{\rho}_0\,
        \op{\mathbb{P}}_s\,\op{\rho}_0\,\op{\mathbb{P}}_s^\dagger 
        \right) \nonumber\\
	    &= (1-p_\text{S}) + p_\text{S}\,\sum_s \sum_{\vect{x},\vect{x}^\prime} 
	        \bra{\vect{x}^\prime,s}\op{\rho}_0\ket{\vect{x},s}
	        \!\bra{\vect{x},s}\op{\rho}_0\ket{\vect{x}^\prime,s} 
	        \nonumber
\end{align}
\begin{align}
    &= (1-p_\text{S}) + p_\text{S}\,\sum_s \sum_{\vect{x},\vect{x}^\prime}
        \left| \bra{\vect{x},s}\op{\rho}_0\ket{\vect{x}^\prime,s} 
        \right|^2 
        \nonumber\\
    &= (1-p_\text{S}) + p_\text{S}\,\sum_s \Bigl(\sum_{\vect{x}}
        \left| \braket{\vect{x},s|E} \right|^2 \Bigr)^2 \,,
\end{align}
where we used the orthogonality of the basis states $\ket{\vect{x},s}$ as well as the purity of the initial state, $\mathrm{tr}(\op{\rho}_0^2) = 1$.
Hence, we obtain
\begin{equation}
    \op{\rho}_1 = \Pi(1)\,\op{\rho}_0 + (1-\Pi(1))\,\op{\tilde{\rho}}_1 ,
\end{equation}
where $\op{\tilde{\rho}}_1$ describes a statistical mixture with no overlap with the initial state, $\mathrm{tr}\left( \ket{E}\! \bra{E}\, \op{\tilde{\rho}}_1 \right) = 0$. 
Assuming that $\ket{E}$ will never be populated by the time evolution of $\op{\tilde{\rho}}_1$,
\begin{equation}
    \mathrm{tr}\left(\ket{E}\!\bra{E}\, {\op{\mathcal{E}}}^{n}(
        \op{\tilde{\rho}}_1) \right) = 0 \quad \forall n > 0 ,
\end{equation}
the probability $\Pi(n)$ to find the walker at time $t=n\,T$ in the initial state is given by
\begin{align}
    \Pi(n) &= \mathrm{tr}\left(\ket{E}\!\bra{E} \op{\rho}_n\right)
    = \mathrm{tr}\left( 
        \op{\rho}_0\, {\op{\mathcal{E}}}^{n-1}( \op{\rho}_1)
      \right)  \nonumber \\
    &=\Pi(1)\, \mathrm{tr}\left(
        \op{\rho}_0\, {\op{\mathcal{E}}}^{n-1}( \op{\rho}_0 )\right) + 
      (1-\Pi(1))\, \mathrm{tr}\left(
        \op{\rho}_0
        {\op{\mathcal{E}}}^{n-1}( \op{\tilde{\rho}}_1)\right)
    \nonumber \\
    &= \Pi(1)^n\, \mathrm{tr}\left( \op{\rho}_0^2 \right) = (1-\gamma_\text{S})^n\,,
\end{align}
where the decay rate $\gamma_\mathrm{S}$ is defined as
\begin{equation}
    \gamma_\text{S} = 1- \Pi(1) = p_\text{S}\, \Biggl[ 
        1 - \sum_s \Bigl(\sum_{\vect{x}}
        \left| \braket{\vect{x},s|E} \right|^2 \Bigr)^2 
        \Biggr].
\end{equation}
For pure position decoherence, one analogously obtains
\begin{equation}
    \Pi(n) = (1-\gamma_\text{P})^n \,,
\end{equation}
where
\begin{equation}
    \gamma_\text{P} = p_\text{P}\, \Biggl[ 
        1 - \sum_{\vect{x}} \Bigl(\sum_{s}
        \left| \braket{\vect{x},s|E} \right|^2 \Bigr)^2 
        \Biggr].
\end{equation}
\end{appendix}

\end{document}